# How Many People Can Simultaneously Move Through a Pedestrian Space?
## The Impact of Complex Flow Situations on the Shape of the Fundamental Diagram

**Dr. Dorine C. Duives (corresponding author)**
Transport & Planning
Delft University of Technology
Stevinweg 1, 2628 CN, Delft, The Netherlands
Email: d.c.duives@tudelft.nl

**Martijn Sparnaaij, MSc.**
Transport & Planning
Delft University of Technology
Stevinweg 1, 2628 CN, Delft, The Netherlands
Email: m.sparnaaij@tudelft.nl

**Dr. Winnie Daamen**
Transport & Planning
Delft University of Technology
Stevinweg 1, 2628 CN, Delft, The Netherlands
Email: w.daamen@tudelft.nl

**Prof. Serge P. Hoogendoorn**
Transport & Planning
Delft University of Technology
Stevinweg 1, 2628 CN, Delft, The Netherlands
Email: s.p.hoogendoorn@tudelft.nl

# How Many People Can Simultaneously Move Through a Pedestrian Space?
## The Impact of Complex Flow Situations on the Shape of the Fundamental Diagram

Duives, D.C., Sparnaaij, M., Daamen, W., Hoogendoorn, S.P.

## ABSTRACT

Pedestrian crowding occurs more frequent. As a result of the increasing pedestrian demand in public space, the limits of pedestrian spaces are of increasing interest. Some research on the maximum demand that can flow through a cross-section has been presented, which mainly features simple movement base cases, low-density situations and/or a homogeneous crowd. Consequently, it is currently unclear to what extent their findings apply to heterogeneous high-density crowds, which are often encountered during real-world scenarios.

The CrowdLimits experiment attempted to reproduce crowd movement dynamics of heterogeneous crowds experiencing higher densities than have been recorded up to this moment. Here, the aim was to study the impact of the three important differences between the current laboratory studies and real-world crowd dynamics in crowded pedestrian spaces simultaneously, namely crowd heterogeneity, high densities movements and (more) complex movement base cases.

This study shows that there are substantial differences in the maximum sustainable flow rate, and the maximum local and global density for distinct movement base cases and flow ratios. Moreover, the results provide evidence of that continuation of flow under very high densities can be recreated under laboratory conditions using a heterogeneous population of pedestrians. Besides, the experimental results indicate that the maximum global flow rate decreases when the scenario becomes more difficult (i.e. bidirectional to intersecting) and the collision avoidance opportunities decrease (i.e. 80-20 to 50-50 flow ratio). Thus, this paper concludes that the maximum flow rate of pedestrian infrastructures decreases with increasing complexity.

## 1. INTRODUCTION

Situations where pedestrian crowding occurs become more frequent. Crowding may occur at busy train stations, or large multi-modal hubs, during mass events in cities, or in busy city centers. As a result of the increasing demand of pedestrian movements in public space, the limits of pedestrian spaces are of increasing interest. In particular, the question 'How much demand can a pedestrian space handle before the crowd movements transition from an ordered effective flow to a chaotic potentially dangerous turbulent flow?' is often raised, in order to determine when the pedestrian infrastructure is running efficiently or is close to the limits of safe and comfortable operations.

Some research on the maximum demand that can flow through a cross-section, often referred to as the capacity of a facility, has been presented (e.g. Fruin et al. (1971), Pretechninski & Milinskii (1978), Seyfried et al. (2005) and Daamen et al. (2010)). These studies often conclude that the capacity of an infrastructure increases approximately linearly with the width of the corridor. Furthermore, the capacity is hypothesized to be dependent on the ratio between flow directions (Kretz et al. 2006). Up to this moment, most empirical studies have either focused on unidirectional movement base cases or studied relatively low-density situations, up to approximately 2 persons per m². However, in large crowds at train stations and event terrains, often far higher densities are (shortly) encountered, up to 5 persons per square meter (Duives (2016) and are often more complex in nature, for instance, bi-directional or intersecting.

Besides that, pedestrians do not always behave 'normal', as they are often in a hurry or perform crossing movements while moving with the flow direction. Similarly, a pedestrian crowd is often very heterogeneous, consisting of young, old, big, small, fast and slow pedestrians. Most recent studies featuring complex movement base cases studied the walking dynamics of a stable homogeneous crowd consisting mainly of young individuals (e.g. Lian et al. 2015 & Cao et al. 2017). As a result, it is currently unclear to what extend their findings apply to heterogeneous high-density crowds.

Consequently, more insights are needed regarding the pedestrian walking dynamics of heterogeneous populations moving in complex movement base cases under high densities. The objective of this study is to determine the impact of movement base cases on the fundamental diagram in case of complex movement base cases featuring a heterogeneous population, where complex refers to the increasing need to adopt collision avoidance strategies to move through a pedestrian infrastructure. Here, we hypothesize that the capacity of the infrastructure is negatively influenced by increasing instabilities in the pedestrian flow during complex movement base cases. In comparison to earlier studies, this study is unique because it studies a heterogeneous participant population.

The paper presents a large pedestrian experiment featuring a population of 130-140 participants from all courses of live is used to study the pedestrian movement dynamics in bidirectional and intersecting movement base cases. Accordingly, fundamental diagrams are derived and the differences between the fundamental diagrams are analyzed. Moreover, some essential characteristics of the fundamental diagrams are quantified, in order to compare the results of this study with previous studies considering homogeneous populations, lower densities or other movement base cases.

The remainder of this paper is organized as follow. First, the state-of-the-art is discussed in section 2. Next, the experimental set-up is presented in section 3. The data gathered during the experiments is elaborated upon in section 4. Section 5 identifies the analysis methodology. Subsequently, section 6 presents and discusses the results. This paper ends with some conclusions and suggestions for future research in section 7.

## 2. BACKGROUND
In recent years, the movement of crowds has been studied quite extensively. Two topics are at the core of these research efforts, namely, 1) the capacity of pedestrian infrastructures and 2) the shape of the fundamental relation between density and flow for pedestrian movements. This section will present a brief overview of the current state-of-the-art concerning these two topics. First, the work on the capacity is presented in section 2.1. Accordingly, section 2.2 discusses the work on the shape of the fundamental relation.

## 2.1. The maximum flow rate in relation to movement base cases

Studies that indicate a maximum flow rate of a pedestrian infrastructure are quite common; some of the most influential have been displayed in TABLE 1. These studies are classified based on the movement base case that has been studied, the flow ratio between movement directions, the conditions at the measurement area and whether or not the pedestrians received instructions with respect to their walking behavior.

The table illustrates that the maximum flow rate varies greatly between these works. The lowest maximum flow rate was established by Virkler & Elayadath (1994), namely 0.97 P/m/s, during a unidirectional flow scenario while a visitor crowd attempted to exit the football stadium of the University of Missouri. The highest flow rate, 3.5 P/m/s, was measured by Löhner et al. (2017) during the circling movement of pilgrims around the Kaaba during the Haji pilgrimage. In general, the table indicates that the more complex the movement base case becomes (i.e. in order of increasing collision avoidance complexity: unidirectional single-file → unidirectional crowd movement → bidirectional → intersecting), the lower the maximum flow rate becomes. However, as the experimental conditions between settings vary greatly, it is difficult to say whether the decrease in the maximum flow rate is solely due to the changing movement scenarios or whether these values are also heavily influenced by the experimental settings.

Only one study has compared the impact of several movement base cases, namely Cao et al. (2017). This study finds only limited differences in the capacity of the maximum flow rate. Consequently, it is difficult to say whether the movement base case influences the maximum flow rate of pedestrian infrastructure.

## 2.2. The shape of the fundamental diagram

A large number of studies featuring the fundamental diagram presented in TABLE 1 illustrates that the shape of the fundamental diagram has been studies, debated and redefined for a long time. Various types of functions have been used to capture the fundamental diagram, from the very simple linear function (e.g. Seyfried et al. (2005)), to the more sophisticated exponential functions (e.g. Weidmann (1993) and Daamen & Hoogendoorn (2003)). A nice summary by Vanumu et al. (2017) illustrates that these differences are not only due to the fact that researchers attempt to use different functions to capture the same variance of data points in one established speed-density graph, but also due to the fact that the underlying data can differ severely between movement scenarios, experimental setups and populations. Secondly, Zhang et al. (2013) illustrated that the metrics used in the fundamental diagram can severely influence the shape of the presented diagrams. Even when taking all of this in consideration, all studies featuring the pedestrian fundamental relation between speed-density and flow-density seem to agree on two properties. First, the fundamental speed-density relation should be a monotonically decreasing function, i.e. $\frac{dV}{d\rho} \leq 0$. Second, the fundamental flow-density relation contains only one (global) maximum value. Besides that, most functions are continuous and do not feature a capacity drop.

However, there is one property of the fundamental speed-density and flow-density relationships that is still heavily debated, namely the shape of the relations when moving towards jam-density. Given that experiments become dangerous when nearing the jam-density, or even end up into a full lock-down of the pedestrian infrastructure, up to this point no laboratory experiments have been performed that can provide details regarding this branch of the pedestrian fundamental diagram. To the authors' knowledge, the only two studies featuring densities >6 P/m$^2$ both studied the Haji Pilgrimage, which should be considered a very special type of pedestrian space (Helbing et al. 2007 & Löhner et al. 2018). These two studies find that also under very high densities (i.e. > 9 P/m$^2$), the pedestrian flow will not completely stop moving forward. Yet, it is unclear whether this is an artefact of the particular case these two studies investigated, or whether this continuation of flow under high densities also occurs in more generic pedestrian infrastructures under pressure, such as transfer hubs and shopping streets.

## 3. EXPERIMENTAL SETUP CROWDLIMITS EXPERIMENT

Studying this type of movements in real-life is difficult due to ethical restrictions. Moreover, one would not be able to record the characteristics of the individuals (e.g. age, gender, length, weight) in the crowd. Thus, Delft University of Technology has set out to find the answer to these two questions by designing a unique pedestrian laboratory experiment – coined CrowdLimits – in which the movement dynamics of pedestrian crowds under very high densities were reproduced.

### 3.1 Aim & operationalization of the CrowdLimits experiment

The CrowdLimits experiment attempted to reproduce crowd movement dynamics of heterogeneous crowds under higher densities than have been recorded up to this moment. Here, the aim was to study the impact of the three major differences between the current laboratory studies and real-world crowd dynamics in crowded pedestrian spaces simultaneously, namely crowd heterogeneity, high densities movements and (more) complex movement base cases. Here, a movement base case refers to one of the flow patterns taken from a comprehensive list of distinctive flow patterns during which a pedestrian an any given time performs only one predominant action.

These three major differences were operationalized using a slightly larger set of input parameters that could be varied one-by-one during a large set of experimental runs. These input parameters are the movement base case, flow ratio, the goal-orientation and predictability of pedestrian movements. The heterogeneity and demand level dynamically varied with time within each experimental run, and as such, were not included in the set of input parameters.

**TABLE 1. Overview of relevant works featuring the fundamental diagram, V=walking speed [m/s], d=pedestrian density[P/m²], f= flow[P/ms], ~ = value derived from graph, - = no information regarding this property provided in the paper, UD-S = uni-directional – single file, UD-C = uni-directional – 2D crowd, UD-B = uni-directional bottleneck, BD = bidirectional, X2 = two-way crossing, X4 = four-way crossing, Xinf = crossing from all directions possible and [1] refers to Buchmuller et al. (2006), and [2] to Cunningham&Cullen (2003).**

| Authors | Scenario | Flow ratio | Conditions | Type of experiment | Capacity [P/m/s] | Formula fundamental relation density - flow |
|---|---|---|---|---|---|---|
| Older (1968) | BD | Variable | Shopping street | Field | 1.26[1] | $q = 1.32k - 0.34k^2$ [1] |
| Fruin (1971) | BD | Variable | Peak-hour flow at commuter bus terminal | Field | 1.46[1] | $q = 1.43k - 0.25k^2$ [1] |
| O'Flaherty & Parkinson (1972) | BD | Variable | Shopping street | Field | 1.29 | $q = 1.53k - 0.49k^2$ |
| Tanabonoon et al. (1986) | BD | Variable | Shopping street | Field | 1.48 | $q = 1.23k - 0.26k^2$ |
| Weidmann (1993) | UD & BD | Variable | Variable | Field | 1.22[1] | $v = v_0 \left( 1 - e^{-1.913\left(\frac{1}{d} - \frac{1}{d_{jam}}\right)} \right)$ |
| Cunningham & Cullen (1993) | BD | Unknown | 'normal' Station traffic | Field | 1.43 | - |
| Virkler & Elayadath (1994) | UD-C | 100-0 | Pedestrian tunnel after University of Missouri football games | Field | 0.97[1] | If $k < 1.07$: $q = 1.01k * \ln(\frac{-k}{4.32})$<br>If $k > 1.07$: $q = 0.61k * \ln(\frac{4.32}{k})$ |
| Sarkar & Janardhan (1997) | BD | Variable | 'normal' Subway traffic | Field | 1.52 | $q = 1.46k - 0.35k^2$ |
| Daamen & Hoogendoorn (2003) | UD-B | 100-0 | Evacuation experiment | Laboratory | ~1.6 | - |
| Seyfried et al. (2005) | UD-S | 100-0 | 'normal' walking motion. | Laboratory | - | $q = 0.36 + 1.06v$ |
| Kretz et al. (2006) | BD | Variable | Laboratory | Laboratory | ~2.3 | - |
| Helbing et al. (2007) | UD-C | 100 -0 | Procession during the Haji | Field | ~ 2.4 | - |
| Chattaraj et al. (2009) | UD-S | 100-0 | Normal | Laboratory | ~1.9 | - |
| Zhang et al. (2012) | UD-C | 100-0 | Normal | Laboratory | ~ 2.4 | - |
| Zhang et al. (2013) | UD-C | 100-0 | Event | Field | 1.8 | - |
| Song et al. (2013) | UD-S | 100-0 | Normal | Laboratory | ~ 2.9 | - |
| Lian et al. (2015) | X4 | 25-25-25-25 | Normal | Laboratory | ~3.3 | - |
| Duives (2016) | UD-C,UD-B, BD, Xinf | Variable | Event | Field | - | - |
| Jin et al. (2017) | UD-C | 100-0 | Normal | Laboratory | ~ 1.7 | - |
| Löhner et al. (2017) | UD-C | 100-0 | Event | Field | ~ 3.9 | - |
| Cao (2017) | UD-C, BD, X2, X4 | 100-0, 50-50, 50-50, 25-25-25-25 | Normal | Laboratory | ~ 1.5, ~ 1.3, ~ 1.5, ~ 1.3 | - |

## 3.2 Movement base-case and site description

This experiment studied two movement base cases that often occur at large-scale events, namely bi-directional flow and intersecting flows under a 90degree angle. FIGURE 1 illustrates the layout of the two movement base cases. These movement base cases were structured using four movable wooden L-shaped panels. Each wall was 2.40 meters high and 3.00 meters long. Jointly, the movable L-shaped panels formed either a long corridor of 2.4 or 1.8 meters wide and 8 meters long, or an intersection of two corridors that had an intersection surface area of 2.4x2.4 or 1.8x1.8 meters. The walls were entirely covered by wooden panels. The construction had no ceiling in order to allow for top-down camera recordings.

In each run of the experiment, the flow distribution over the entrances was varied. The entrance flow was either distributed 50-50 over both entrances or split in a major flow (80%) and a minor flow (20%). These two particular cases are adopted because the work of Kretz et al. (2006) illustrates that these two flow ratio scenarios are the most advantageous (50-50) and disadvantageous (80-20) flow ratio scenarios with respect to the resulting capacity. Please note, 80%-20% are the designed flow rates. In reality, the major flow varied between 70 and 90%, while the minor flow varied between 10 and 30%.

FIGURE 1a and c show the movement patterns in the 50-50 case and figure 1b and d the movement pattern in the 80-20 case. In the 80-20 case, all participants were asked to follow the major flow direction (black arrows in figure 1b and figure 1d). The smaller queue was replenished with participants that were randomly picked out of the major flow by two staff members. This was done in order to ensure that not always the same persons would end up in the minor flow, thus ensuring that the heterogeneity of the major and minor flow dynamically changed over time within each run. At the highest flow rates, in some runs difficulties arose with the replenishments of the queues at the entrances, due to a strong decrease of the outflow rate of the construction. In those cases, the experiment was stopped ahead of time, as no recuperation was to be expected within the runtime of the experiment.

## 3.3 Experimental set-up and scenarios

During each run, the flow rate was increased every minute. TABLE  illustrates the respective flow rates during the experiments. The minimum designed flow rate is set to be the maximum flow rate at which continuous flow could still be guaranteed. This flow rate is determined using experience from earlier works by Zhang et al. 2014 and Wong et al. (2010) in combination with a pilot run that was performed during the days of the experiments. The maximum flow rate was determined by the minimum walking time of the participants. The inflows were controlled using stop-and-go signals at all entrances.

Each of the participants received an assignment at registration, which was aimed to change the goal-orientation and predictability of the movement behavior of a part of the participants in order to stimulate the emergence of more 'chaotic' crowd movement dynamics. This was aimed to destabilize the generally stable crowd movement dynamics, thus increasing the 'complexity' of the collision avoidance behavior. Each participant only knew their own assignment as they were asked to keep their assignment hidden from the other participants. This in order to ensure that other participants could not 'help' other participants to perform their assignment. In total, three distinct assignments were handed out: A) *No assignment, B) Crossing* - aimed to increase goal-orientation and thus pressure differences in the crowd, C) *Intersecting* - designed to decrease the predictability of the crowd movements.

The participants with assignment A, which totaled approximately 80% of the participants, were asked to always perform their assignment. The participants with assignments B and C, which each totaled 10% of the total number of participants in size, were asked only to perform their assignment at the moment the stop-and-go signal indicated the letter of their assignment. Participants were asked and repeatedly reminded to perform their assignment while they were located inside the construction. On both days, participants had to be reminded of their assignment, as the assignment is very counterintuitive with respect to their natural behavior.

During two days, all combinations of the two movement base cases, two flow ratios and three assignments were tested twice. This results in 24 experiments; see TABLE 3 for the order in which the combinations were tested. For this paper, the four base scenarios are studied, namely 1) Bidirectional 50-50 without assignment; 2) Bidirectional 80-20 without assignment; 3) Intersecting 50-50 without assignment; 4) Intersecting 80-20 without assignment.

## 4. DATA COLLECTION AND DATA EXTRACTION

The movement dynamics of each individual in the crowd was captured using a set of video cameras attached to the ceiling of an examination hall facing top-down at the height of 8 meters. The center camera, an 8MP camera, was directed at the center of the infrastructure and captured the approach and exiting of all participants. Two other Logitech C930E HD Pro Webcam cameras 1080p 21MP, hereafter identified as 'c930 left' and 'c930 right' were directed at both approaches. FIGURE 2a and FIGURE 2b visualize the field of view of the three cameras. These two cameras also captured the center of the infrastructure but did not capture the exit of the infrastructure.

The participants of the experiment were, on average, $29.2 \pm 13.0$ years old, were $1.77 \pm 0.10$ meters long and weighted $72.5 \pm 14.4$ kilograms. Approximately 74% considered themselves Dutch citizens. Each of the participants was provided with a red cap and a white t-shirt that covered the participants' outer clothing. The cap was marked with a white center point and a white barcode (see FIGURE 3a). The center point is used to identify the position of the head in all frames of the videos. The reader is referred to FIGURE 3b for an example of the identification of the center point of the participants head.

A combination of various computer vision algorithms predefined in Matlab, such as blob detection in RGB, binary erosion and morphological closing, is used to automatically detect the position of the cap of all the participants in all frames of the three selected videos. Accordingly, a combination of a trained neural network, Kalman Filtering algorithm and a multi-object tracking algorithm (i.e. Munkres (1957)) are adopted to track the movements. The automatic computer vision detection and tracking procedure identify 90% of the trajectories correctly. To ensure 100% correct detection of all trajectories, all trajectories have been manually checked and corrected if necessary. An example of the result of the total procedure is depicted in FIGURE 4.

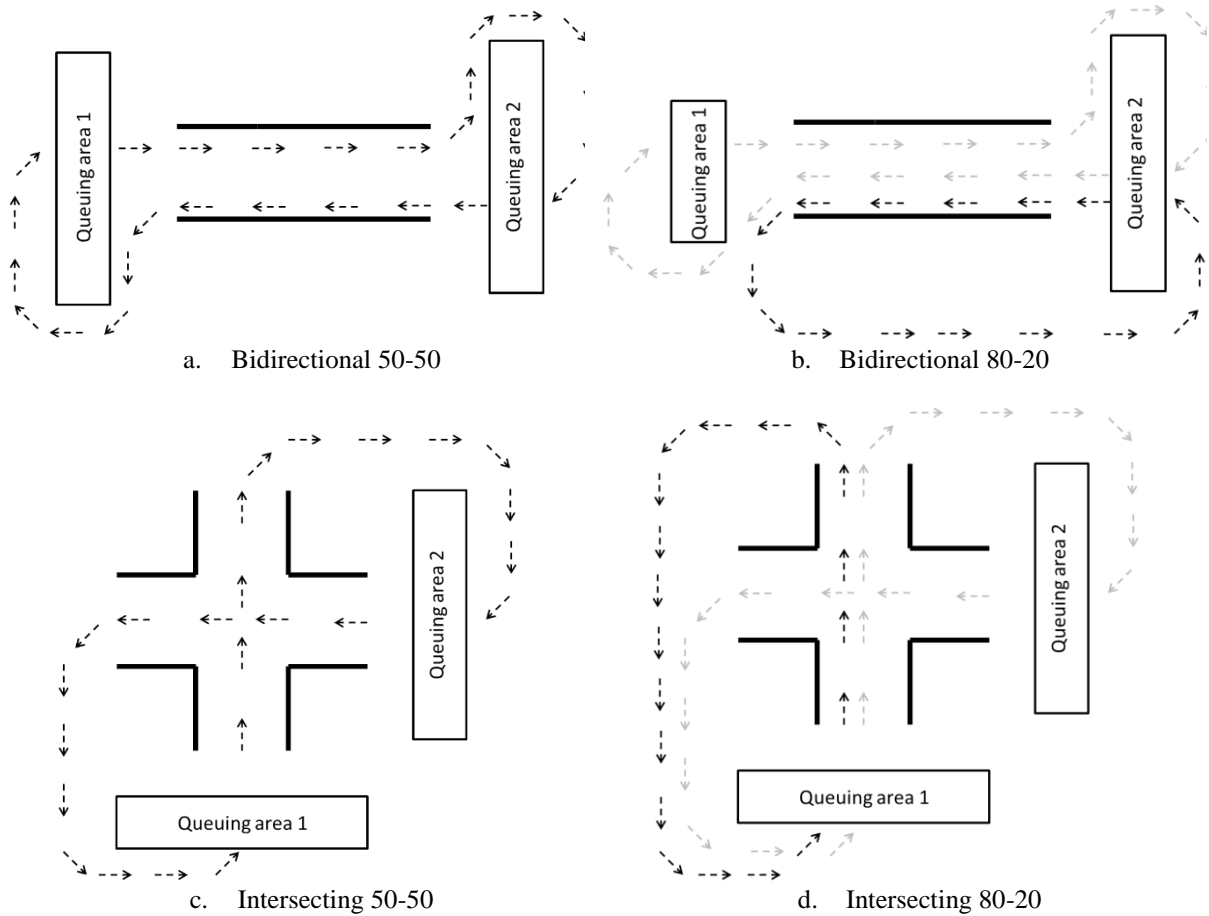

a. Bidirectional 50-50      b. Bidirectional 80-20

c. Intersecting 50-50      d. Intersecting 80-20

**FIGURE 1. Movement scenarios for June 5[th] (A & B) and June 6[th] (C & D), where the black arrows indicate the movement direction of the major flow and the grey arrows of the minor flow.**

**TABLE 2. Distribution of flowrates in pedestrians per second per direction over the scenarios**

| Time [min:sec] | Bi-directional 50-50 | Bi-directional Minor-major | Intersecting 50-50 | Intersecting Minor-major |
|---|---|---|---|---|
| 0:00 – 1:00 | 1.68 | 0.67 / 2.69 | 1.91 | 0.77 / 2.05 |
| 1:00 – 2:00 | 1.92 | 0.77 / 3.07 | 2.16 | 0.86 / 2.30 |
| 2:00 – 3:00 | 2.16 | 0.86 / 3.46 | 2.40 | 0.96 / 2.64 |
| 3:00 – 4:00 | 2.64 | 1.06 / 3.84 | 2.64 | 1.06 / 2.82 |
| 4:00 – 5:00 | 2.88 | 1.15 / 4.22 | 2.88 | 1.15 / 3.07 |

**TABLE 3. Distribution of scenarios over the two experimental days**

| Run no. | Day 1 | | | Day 2 | | |
|---|---|---|---|---|---|---|
| | Movement base case | Flow distribution | Assignment | Movement base case | Flow distribution | Assignment |
| 1 | Bidirectional | 50-50 | No | Intersecting | 50-50 | No |
| 2 | Bidirectional | 50-50 | No | Intersecting | 50-50 | No |
| 3 | Bidirectional | 50-50 | No | Intersecting | 50-50 | Crossing |
| 4 | Bidirectional | 50-50 | Crossing | Intersecting | 50-50 | Fast walk |
| 5 | Bidirectional | 50-50 | Fast walk | Intersecting | 50-50 | Crossing |
| 6 | Bidirectional | 50-50 | Crossing | Intersecting | 50-50 | Fast walk |
| 7 | Bidirectional | 90-10 | Fast walk | Intersecting | 80-20 | No |
| 8 | Bidirectional | 80-20 | No | Intersecting | 80-20 | No |
| 9 | Bidirectional | 80-20 | Crossing | Intersecting | 80-20 | Crossing |
| 10 | Bidirectional | 80-20 | Fast walk | Intersecting | 80-20 | Fast walk |
| 11 | Bidirectional | 80-20 | Crossing | Intersecting | 80-20 | Crossing |
| 12 | Bidirectional | 80-20 | Fast walk | Intersecting | 80-20 | Fast walk |

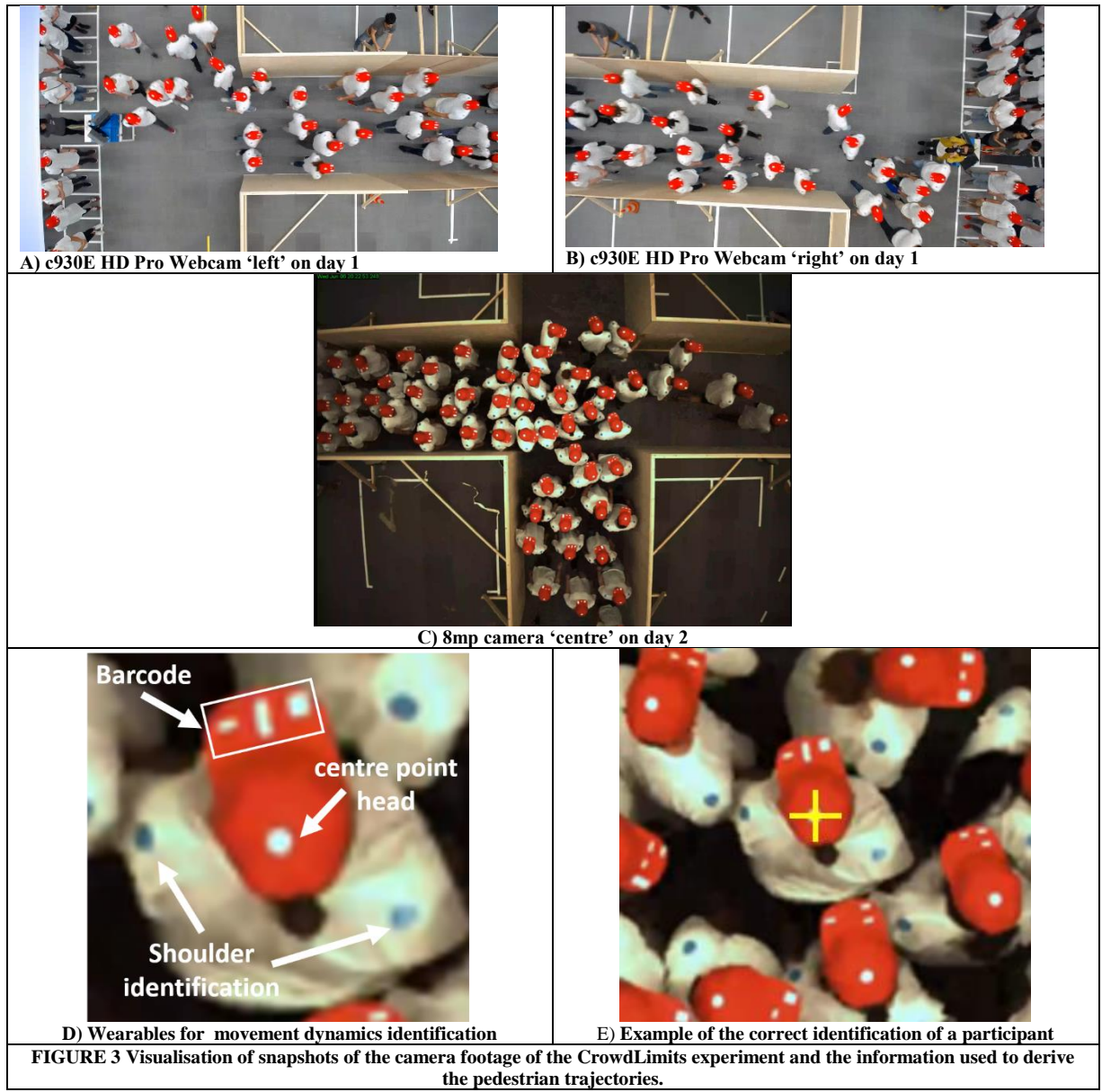

**A) c930E HD Pro Webcam 'left' on day 1**

**B) c930E HD Pro Webcam 'right' on day 1**

**C) 8mp camera 'centre' on day 2**

**D) Wearables for movement dynamics identification**

**E) Example of the correct identification of a participant**

**FIGURE 3 Visualisation of snapshots of the camera footage of the CrowdLimits experiment and the information used to derive the pedestrian trajectories.**

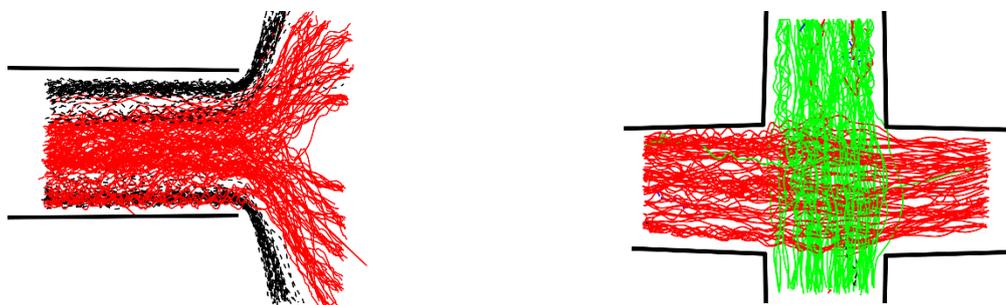

**FIGURE 4 Example of a randomly selected cleaned set of 200 trajectories for the bi-directional movement scenario and the intersecting movement scenario, where the red lines identify the trajectories of participant coming from the right, the dashed black lines participants are coming from the left and the dashed green lines participants moving from bottom to top.**

## 5.   QUANTIFYING THE IMPACT OF MOVEMENT BASE CASES

This section presents a mathematical description of the variables that will be used to quantify the differences in the fundamental diagrams. These variables will be used in section 6 to analyze the data from the CrowdLimits experiment. First, the derivation of the pedestrian flow variables from the pedestrian trajectory data is described. Accordingly, the methods to quantify the differences in the fundamental diagram are provided.

### 5.1 Derivation of pedestrian flow variables

The fundamental diagram displays the relation between the effective speed and density. In recent years, there has been some discussion considering the correct and most valid manner to compute these properties for pedestrian flows. The Voronoi method to compute the speed and density will also be used in this paper, given that Steffen & Seyfried (2009) and Zhang et al. (2011) illustrate that the Voronoi method is currently one of the most reliable methods, and this method is often adopted by other studies. The density experienced by a participant is determined as the inverse of the area of the Voronoi Cell, which is a decomposition of the metric space where each point in space is assigned to the pedestrian nearest to the point (eq. 1), where $A_i$ is the area of the Voronoi cell corresponding to participant $i$.

To quantify the global density $\rho(t)$ the method proposed by Steffen & Seyfried (2009) is adopted, hereafter coined the global Voronoi density. In essence, the number of pedestrians located inside the measurement area N divided by the summation of the area of the Voronoi cells occupied by those pedestrians. As a result of this definition, the exact area taken into account can vary over time and can be bigger or smaller than the area of the measurement area.

$$\rho_i(t) = \frac{1}{A_i} \tag{1}$$

$$\rho(t) = \frac{\int_A p(\vec{x})\,d\vec{x}}{|A_i|}, \ \forall \ i \ \in A \tag{2}$$

The speed is defined as the effective speed in the direction of movement, which is operationalized as the projection of the velocity vector on the vector pointing in the general heading of the corridor the pedestrian traverses. See eq. 3 for the mathematical formulation of the velocity, where $\vec{X}_{i,x}(t)$ is either the x-coordinate or the y-coordinate of the position of participant $i$ at time $t$ and $\Delta t$ the time period between five consecutive video frames, depending on the direction of movement. Here, the effective speed of all participants walking should always be larger or equal to 0 m/s.

The global speed $v(t)$ of area at time t is defined as the average of the walking speeds of all pedestrians currently residing inside the measurement area $A$. Please note, that the speed is the magnitude of the velocity, and as such, is a scalar instead of a vector.

$$v_i(t) = \left\| \frac{\vec{X}_{i,x}(t+\Delta t) - \vec{X}_{i,x}(t)}{\Delta t} \right\| \tag{3}$$

$$v(t) = \frac{\sum_i v_i(t)}{N}, \ \forall \ i \ \in A \tag{4}$$

The flowrate is computed as the multiplication of the speed and the density. That is, the local (or rather, 'individual') flow rate measurement is the multiplication of the effective speed $v_i(t)$ (in the x-direction) times the Voronoi density $\rho_i(t)$; the global flow rate measurement is the multiplication of the average effective speed $v(t)$ times the global Voronoi density $\rho(t)$. Please note that the flow rate (in the way we defined it) is a scalar and has no direction. Thus, higher speeds at a similar density result in a higher flowrate, irrespective of the direction in which the pedestrians are moving as the speeds cannot cancel each other out.

$$q_i(t) = v_i(t) \cdot \rho_i(t) \tag{5}$$
$$q(t) = v(t) \cdot \rho(t) \tag{6}$$

## 5.2 Quantifying the shape of the fundamental diagram

Accordingly, the fundamental diagrams can be determined. It can be challenging to determine the differences between the shapes of the fundamental diagrams. Thus, each of the following points is determined:

$\rho_{max}$: The maximum density measured during the run (99th percentile)

$q_{max}$: The maximum flow rate measured during the run (99th percentile)

$v_0$: The maximum effective speed measured during the run (99th percentile)

## 6. RESULTS

In this section the fundamental diagrams resulting from the four distinct 'basic' scenarios of the CrowdLimits experiment and discusses the differences in the shape of the fundamental diagrams. First, a visual comparison is made of both the local fundamental diagrams as well as the global fundamental diagrams.

## 6.1 Local pedestrian fundamental diagram

FIGURE 5 presents the four resulting fundamental diagrams, in which the local flow rate (i.e. the multiplication of the speed of a pedestrian multiplied by the inverse of the area of the Voronoi cell of a pedestrian) is related to the local density for a given point in time. Most surprisingly, none of the diagrams have the distinctive mountain-like shape of a fundamental diagram, which suggests one of two things, A) the maximum flow rate of the pedestrian infrastructure was not reached in any of the four experimental runs, or B) at the level of an individual the flow rate is fluctuating highly as the result of the lacking temporal dimension. That is, pedestrians might briefly accept high densities while retaining their high walking velocity in order to gain a long-term advantage. The results of the local fundamental diagram cannot be used to identify which of these two lines of reasoning is correct.

Secondly, the gradient of the maximum flow rate of the intersecting scenario is steeper than the gradient of the bidirectional cases (see FIGURE 5e). This means that at the same density, the pedestrians tend to walk faster in the intersecting scenario than in the bidirectional case. In addition, this finding points towards a more long-term strategy of the pedestrians during the intersecting scenario. The question is, if one takes into account the temporal dimension of density, i.e. how long pedestrians experience a certain density, whether this finding will still hold for all time scales. More research in this direction is encouraged.

Thirdly, the maximum density measured locally is higher in the bidirectional scenarios, as can be seen in TABLE 4. The more equal the flows are divided, the higher the maximum density becomes. The maximum flowrate and maximum speed seem to follow another pattern, which is more in line with the steepness of gradient of the flow-density diagrams.

Lastly, large differences in the maximum flow rate at a given density for the bidirectional scenarios are found, while these are almost non-existent between the two intersecting scenarios. The authors hypothesize that this is the result of a difference in the opportunities to solve pending collisions. In the intersecting scenarios, the opportunities to solve collisions are similar when meeting a large flow of pedestrians who want to orthogonally cross your flow. In the bidirectional case, the opportunities severely diminish if the counter-flow is relatively large.

## 6.2 Global pedestrian fundamental diagram

The global fundamental diagram is visualized in FIGURE 6, which depicts the relation between the effective flow rate and the average density. This diagram shows that the local differences in the flow rate cancel out almost completely. As a result, the distinctive shape of the fundamental diagram is found in at least two of the four scenarios, namely the bidirectional 80-20 and intersecting 50-50 case. In the bidirectional 50-50 scenario, the onset of the decrease of the maximum flow rate is found, but not the expected steep decline at higher densities. In the intersecting 80-20 case, no decrease of the slope of the maximum flow rate with the density is seen.

Interestingly, the density at which the gradient of the normalized maximum flow rate starts to decline or becomes 0 differs between scenarios. The scenario with the lowest density at the capacity point is the intersecting 50-50 scenario, at a density of approximately 2 P/m$^2$. The scenario that last reaches the maximum flow rate is the bi-directional 50-50 case, at a density of almost 8 P/m$^2$. Based on the previous work by, amongst others, Lian et al. (2015) and Cao et al. (2017), these differences are quite unexpected. This finding suggests that relatively more space is needed to efficiently move in an intersecting scenario than, for instance, bi-directional flows.

Thirdly, similar to the local fundamental diagrams, the density was found to be a lot higher in both bidirectional scenarios than in both intersecting cases. One possible explanation for this contrast is the difference in the area in which the two flows interact. In the bi-directional case, the flows interact over the entire length of the corridor. Consequently, density can build up over time and space. In the intersecting scenarios, the flows interact only at the intersection; as such, the density can dissipate a lot better.

The last major difference between the diagrams is the maximum normalized flow rate (see TABLE 4). Contrary to the local fundamental diagram, the largest flow rate normalized is sustained in the bidirectional scenarios, in particular, the 80-20 scenario. Moreover, the maximum normalized flow rate is larger in both 80-20 scenarios, in comparison to the 50-50 scenarios, which suggest that in the 50-50 scenarios more friction occurs.

**TABLE 4 95th percentile of the speed, density and flow computed using the local and global interpretation of the flow variables**

| | Bidir 50-50 | Bidir 80-20 | Intersect 50-50 | Intersect 80-20 |
|---|---|---|---|---|
| | **Local interpretation of flow variables** | | | |
| $\rho_{max}$ | 8.74 | 7.19 | 4.45 | 3.27 |
| $q_{max}$ | 2.01 | 2.85 | 3.14 | 2.64 |
| $v_{max}$ | 0.34 | 0.67 | 1.21 | 1.07 |
| | **Global interpretation of flow variables** | | | |
| $\rho_{max}$ | 7.00 | 7.25 | 2.92 | 2.67 |
| $q_{max}$ | 0.65 | 0.86 | 0.23 | 0.54 |
| $v_{max}$ | 0.35 | 0.58 | 1.12 | 0.98 |

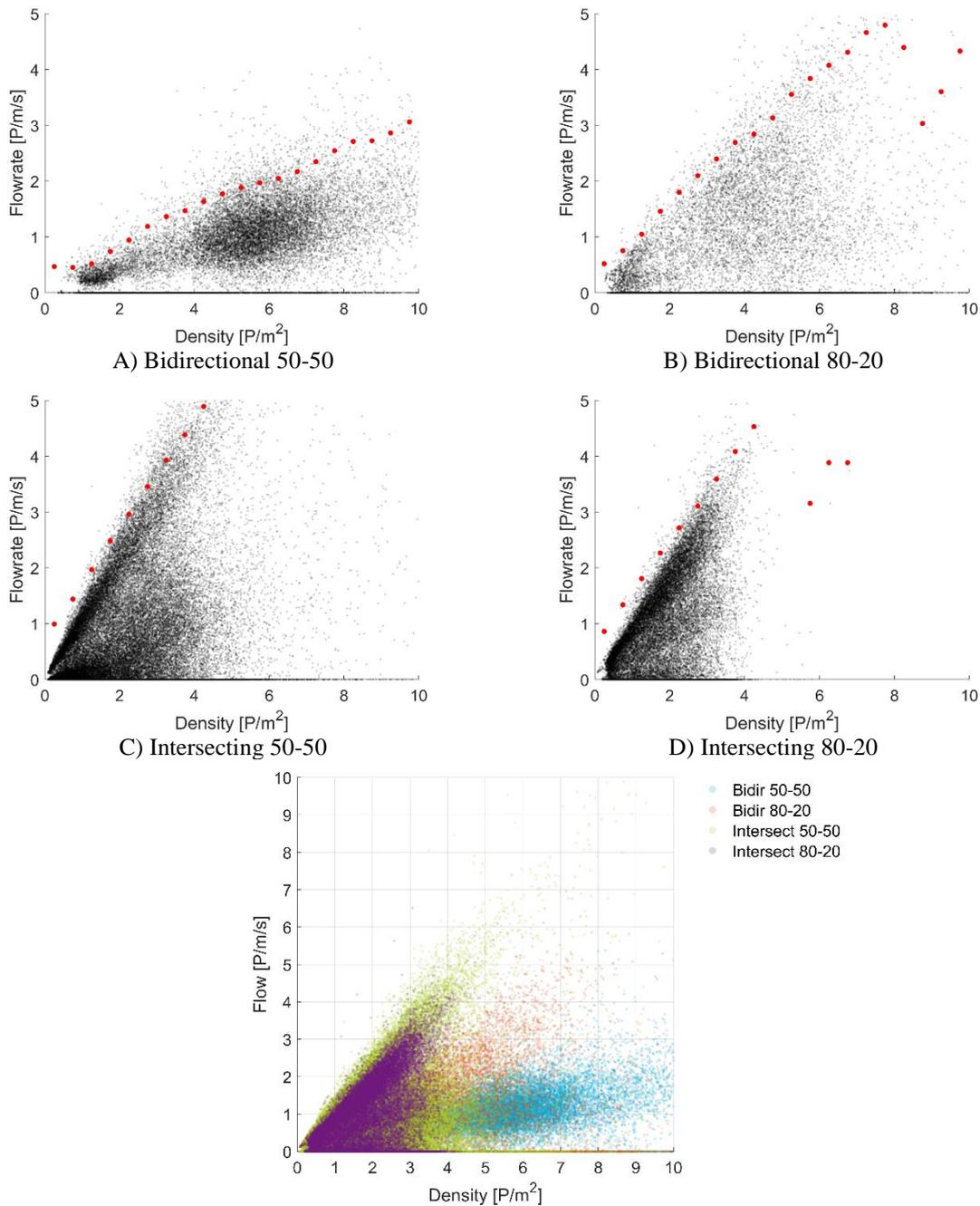

A) Bidirectional 50-50

B) Bidirectional 80-20

C) Intersecting 50-50

D) Intersecting 80-20

E) Comparison of the local fundamental diagrams

**FIGURE 5 Visualization of the local fundamental diagrams for the four scenarios (A-D), and E) a comparison of the fundamental diagrams**

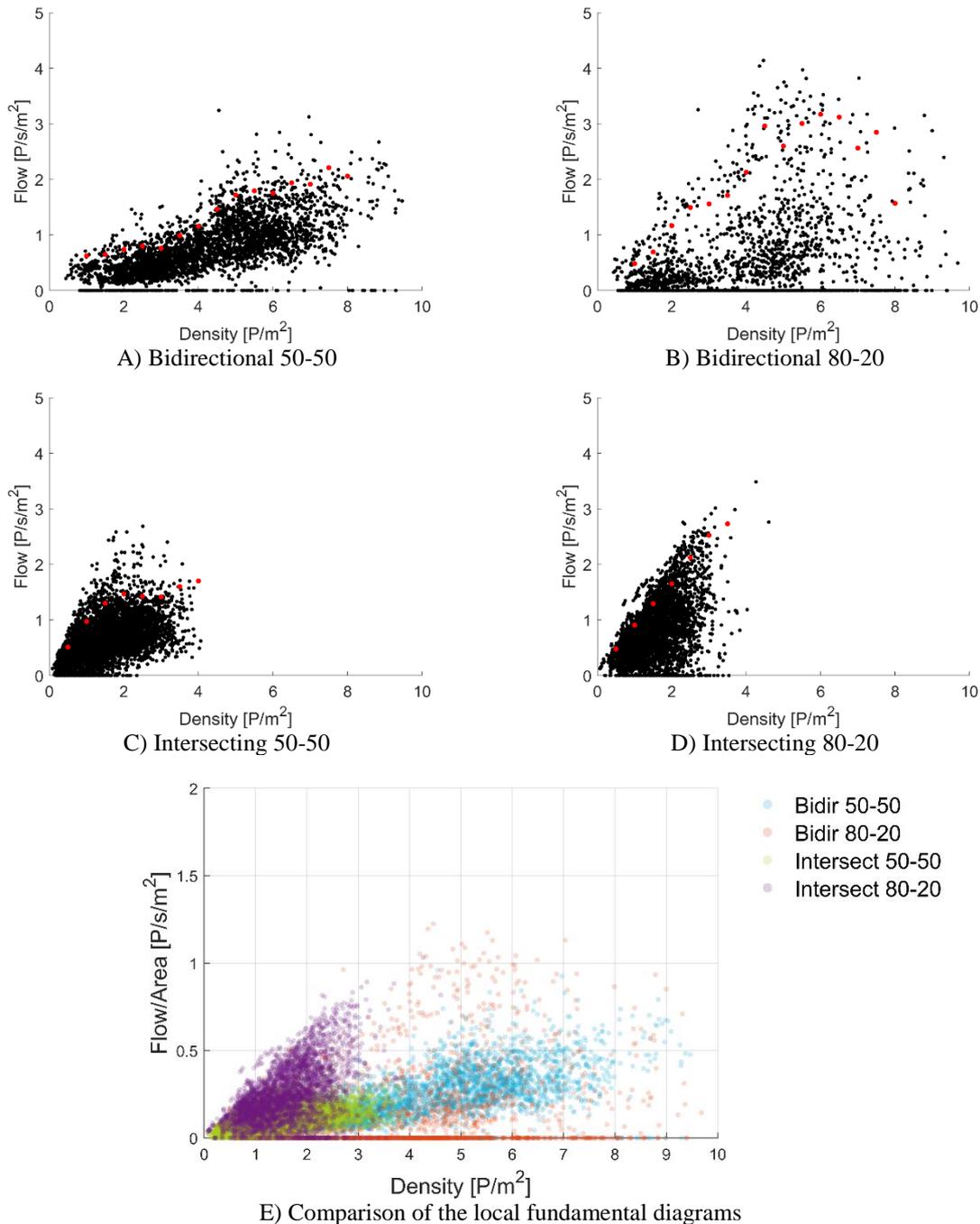

E) Comparison of the local fundamental diagrams

**FIGURE 6. Visualization of the local fundamental diagrams for the four scenarios (A-D), and E) a comparison of the fundamental diagrams**

## 7. CONCLUSION & FUTURE WORK

This research has studied the shape of the pedestrian fundamental diagram from a local and global perspective. Data from a large laboratory study coined CrowdLimits featuring a heterogeneous crowd were used to study the impact of differences in the movement base case and flow ratio. Trajectory data was derived from video recordings, which were used to visually and quantitatively compare the resulting fundamental diagrams.

The results in this paper are only partly in line with previous studies. That is, this study provides evidence of that continuation of flow under very high densities, which was first mentioned by Helbing et al. (2007) and afterwards recreated by Lian et al. (2015) using an Asian homogeneous student population, is not an artefact, nor culture-specific, but can also be recreated under laboratory conditions using a European heterogeneous population of pedestrians. At the same time, Cao et al. (2017) concluded that there are only small differences in the fundamental diagrams featuring distinct movement base cases. The results of the CrowdLimits experiment point in the opposite direction, showing substantial differences in the maximum sustainable flow rate, and the maximum local and global density.

We postulated the hypothesis that the capacity, i.e. maximum flow rate, of the infrastructure is negatively influenced by increasing instabilities in the pedestrian flow, which are created by the introduction of more complex movement base cases. The results support this hypothesis, showing that the maximum normalized global flow rate decreases when the scenario becomes more difficult (i.e. bidirectional to intersecting) and the collision avoidance opportunities decrease (i.e. 80-20 to 50-50 flow ratio). Thus, this paper concludes that the capacity of pedestrian infrastructures decreases with increasing complexity.

This paper presented preliminary results of the CrowdLimits experiment and generated several new questions. Directions of future work include studying the impact of the time period during which pedestrians experience a certain density on their walking speed and researching the impact of the assignments, and thus goal-orientation and unpredictability of crowd movements, on the shape fundamental diagram.

**ACKNOWLEDGEMENTS**

The research presented in this paper is part of the research program "Allegro: Unravelling slow mode travelling and traffic: with innovative data to a new transportation and traffic theory for pedestrians and bicycles" (ERC Grant Agreement no. 669792), a Horizon 2020 project which is funded by the European Research Council.

**AUTHOR CONTRIBUTIONS**

The authors confirm contribution to the paper as follows: study conception and design: D. Duives, M. Sparnaaij; data collection: D. Duives, M. Sparnaaij, W. Daamen; analysis and interpretation of results: D. Duives; draft manuscript preparation: D. Duives, M. Sparnaaij, W. Daamen, S.P. Hoogendoorn. All authors reviewed the results and approved the final version of the manuscript.